\documentclass[11pt]{article}

\usepackage[margin=1in]{geometry}
\usepackage{hyperref}
\usepackage{graphicx}

\title{Autonomous Editorial Systems and Computational Investigation with Artificial Intelligence}
\author{
Ahmed Banafea \\
Betron Labs LLC \\
}
\date{\today}

\begin{document}
\maketitle

\begin{abstract}
Understanding the news involves more than reporting individual facts. It requires determining which stories matter, how information from different sources connects, and what conclusions emerge when reporting is examined collectively over time. As the public news record grows in volume, continuity, and fragmentation, these activities increasingly operate at a scale where system-level approaches become effective.

This paper describes a class of artificial intelligence systems that perform two core functions automatically: autonomous editorial organization and computational investigation. These systems use large language models for semantic interpretation, embedding-based representations for similarity and alignment, and structured storage for maintaining persistent editorial state. Together, these components enable large volumes of reporting to be organized into long-lived story structures and examined through systematic comparison across sources and time.

The framework presented here emerged directly from the design and operation of World Pulse Now (WPN News)\footnote{\url{https://www.worldpulsenow.com}}, an agentic editorial system currently operating in production. While WPN News is publicly positioned around autonomous editorial organization, its underlying architecture supports a broader capability: using artificial intelligence to investigate the public record through large-scale comparison and inference.

By treating stories as persistent computational objects, separating editorial organization from investigative analysis, and orchestrating artificial intelligence components deterministically, the system transforms informational scale from a limitation into a source of insight.
\end{abstract}

\section{Introduction}

The scale of contemporary public news has grown beyond what can be coherently interpreted through manual editorial processes alone. Major global events routinely generate hundreds or thousands of articles across different outlets, regions, and political contexts, often within short periods of time. While this abundance increases access to information, it also creates a coordination problem: understanding how large volumes of reporting collectively describe evolving events.

Making sense of the news requires more than reading individual articles. It involves identifying which reports refer to the same underlying event, understanding how accounts differ across sources, tracking how narratives change over time, and recognizing when new developments meaningfully alter the overall picture. Traditionally, these tasks have been carried out by human editors and investigators, operating under significant attention and time constraints.

As reporting volume and continuity increase, these activities begin to resemble computational problems. Prior work in information retrieval and information extraction has treated large-scale text streams as event-oriented systems that require continuous detection, monitoring, and comparison across documents and time \cite{chen2020history, xiang2019survey}. Recent advances in artificial intelligence—particularly large language models and representation learning—make it possible to approach these challenges through system design rather than episodic human judgment.

This paper explores how artificial intelligence can be used to implement two core capabilities directly as system functions: autonomous editorial organization and computational investigation. Rather than assisting humans on a case-by-case basis, the systems described here continuously read, compare, and aggregate reporting at scale, across thousands of stories simultaneously. Editorial structure and investigative insight emerge from the system’s operation itself.

The ideas presented in this paper emerged from the design and deployment of World Pulse Now (WPN News), an agentic system that continuously organizes and analyzes the public news record. While WPN News is publicly focused on autonomous editorial organization, building the system revealed a broader insight: given sufficient independent reporting, artificial intelligence can not only organize the news into consistent, evolving story structures, but also investigate the public record through structured comparison and inference.

\section{Editorial Systems and Investigation at Scale}

News production has traditionally been organized around distinct human roles. Reporters gather information and describe events. Editors decide how stories are grouped, prioritized, and explained. Investigators look across sources and time to identify patterns, timelines, and broader implications. This division of labor works well at small scale.

At large scale, however, it runs into human limits. No editor can continuously track thousands of related articles across extended periods. No investigative team can systematically compare every claim made about a developing event across dozens of sources. As reporting volume grows, the bottleneck is no longer access to information, but the ability to synthesize it.

Much investigative insight arises from synthesis rather than the discovery of new facts. When multiple sources provide partial, evolving, or conflicting accounts, new understanding emerges through comparison. Prior computational work on event detection, tracking, and story evolution in streaming text corpora has already demonstrated how such dynamics can be modeled at scale \cite{ahmed2011unified, rose2009describing, peng2021streaming}.

Artificial intelligence is particularly well suited to this setting. AI systems can read continuously, maintain structured memory across large corpora, and apply consistent comparison logic over long time horizons. As a result, both editorial organization and investigation naturally shift toward persistent computational processes rather than episodic human intervention.

This shift enables a different way of thinking about editorial work and investigation. Instead of being treated as manual tasks performed intermittently, they can be designed as system-level operations: continuously running processes that organize, compare, and interpret reporting at scale.

\section{Autonomous Editorial Systems}

Editorial work consists of a set of concrete operations that can be described independently of newsroom roles. These operations include identifying which articles describe the same event, determining whether information is new, tracking how an event changes over time, assessing relative importance, and generating explanations that reflect the current state of affairs.

Autonomous editorial systems use artificial intelligence to carry out these operations directly. Such systems ingest large volumes of reporting, interpret article content, and group related items into organized event structures. As new articles arrive, the system updates its representation of each event and adjusts its editorial outputs accordingly.

A defining characteristic of autonomous editorial systems is that stories are treated as persistent, stateful structures rather than transient collections of articles. Articles serve as inputs, while stories accumulate context, history, and structure over time. This approach aligns with earlier system designs in which documents are treated as inputs to evolving narrative or event models \cite{ahmed2011unified, peng2021streaming}.

Editorial outcomes are produced by the system itself. Humans define system goals, constraints, and operating parameters, but they do not select individual headlines, summaries, or rankings at publication time. Editorial organization becomes a continuous computational process embedded directly in the system’s architecture.

Autonomous editorial systems differ from personalization or recommendation systems, which are optimized around user behavior. Instead, they are designed to support narrative coherence, continuity, and explanatory clarity across the public record.

\begin{figure}[h]
\centering
\includegraphics[width=\textwidth]{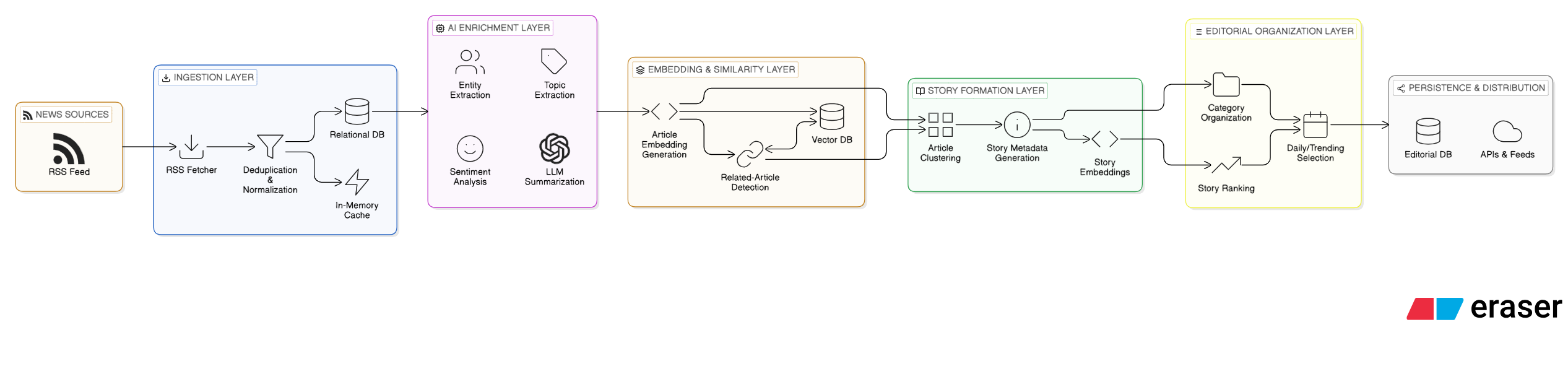}
\caption{End-to-end architecture of the autonomous editorial system, from article ingestion to editorial outputs.}
\label{fig:architecture}
\end{figure}

Figure~\ref{fig:architecture} illustrates the overall system architecture, showing how articles move from ingestion through enrichment and embedding, into persistent story structures that support editorial organization.

\section{Computational Investigation}

Investigation can be understood as the process of discovering what emerges when many pieces of information are examined together. Computational investigation applies this idea at scale using artificial intelligence.

Rather than collecting new evidence, the system examines relationships among existing reports. It compares how different sources describe the same event, tracks how claims evolve over time, and identifies patterns such as convergence, divergence, delayed confirmation, and narrative shifts. Prior research has shown the value of this longitudinal, cross-source approach for understanding story evolution and event dynamics \cite{rose2009describing, mele2019event}.

When independent sources align on a claim, interpretive confidence strengthens. When accounts diverge or change, transitions become visible. These signals arise from systematic comparison across large volumes of reporting rather than from isolated reading of individual articles.

Artificial intelligence enables this process by maintaining structured representations of claims and monitoring them across sources and time. Investigation becomes an inferential activity in which understanding emerges from structure, repetition, and change.

Computational investigation operates alongside autonomous editorial systems. Editorial organization summarizes what is currently known, while computational investigation explores what the broader public record reveals when examined as a whole.

\section{The Public Record as a Computational Substrate}

For autonomous editorial systems and computational investigation, the public news record is treated as structured input rather than a collection of isolated documents. Each article contains elements such as claims, references, entities, sources, and timestamps that can be explicitly extracted and represented \cite{xiang2019survey}.

When aggregated, these elements form a rich computational substrate. Claims connect across sources, stories persist over time, and narrative trajectories become observable as reporting accumulates. Prior work in narrative visualization and temporal analytics has emphasized the importance of persistent representations for understanding how information evolves \cite{fisher2008narratives}.

Artificial intelligence systems operate on this substrate by identifying relationships among claims and updating those relationships continuously as new reporting arrives. Independence among sources strengthens signals, while divergence highlights uncertainty or transition. Temporal depth allows patterns to be examined across longer horizons rather than at a single moment in time.

\begin{figure}[h]
\centering
\includegraphics[width=\textwidth]{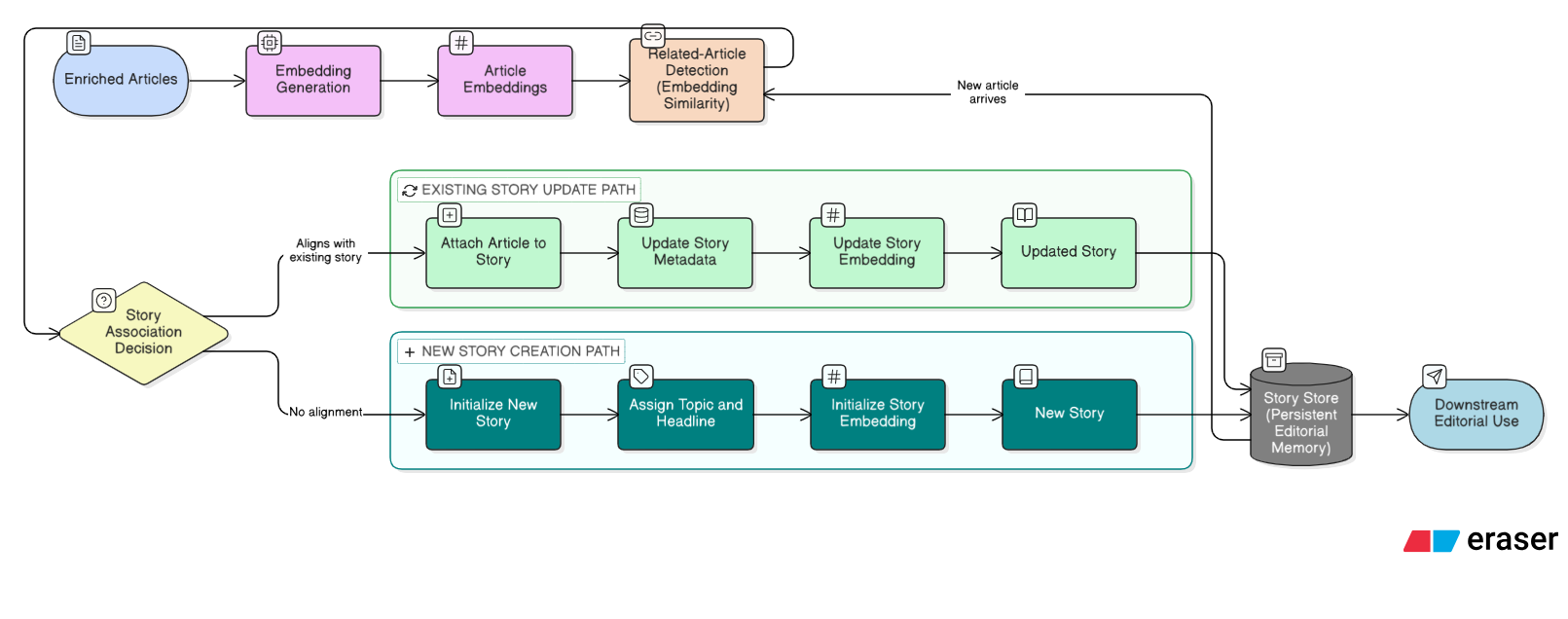}
\caption{Embedding-based story formation and persistent editorial memory. Incoming articles are associated with existing stories or initiate new ones, allowing stories to evolve over time.}
\label{fig:story_memory}
\end{figure}

Because stories persist and accumulate context, each additional article increases the system’s informational value. Figure~\ref{fig:story_memory} illustrates how articles are organized into persistent story structures that build context across sources and time.

\section{System Architecture}

The system architecture enables autonomous editorial organization and computational investigation through a structured pipeline that operates continuously over incoming reporting. It treats stories as persistent state, separates editorial organization from investigative analysis, and orchestrates artificial intelligence components deterministically.

\begin{figure}[h]
\centering
\includegraphics[width=\textwidth]{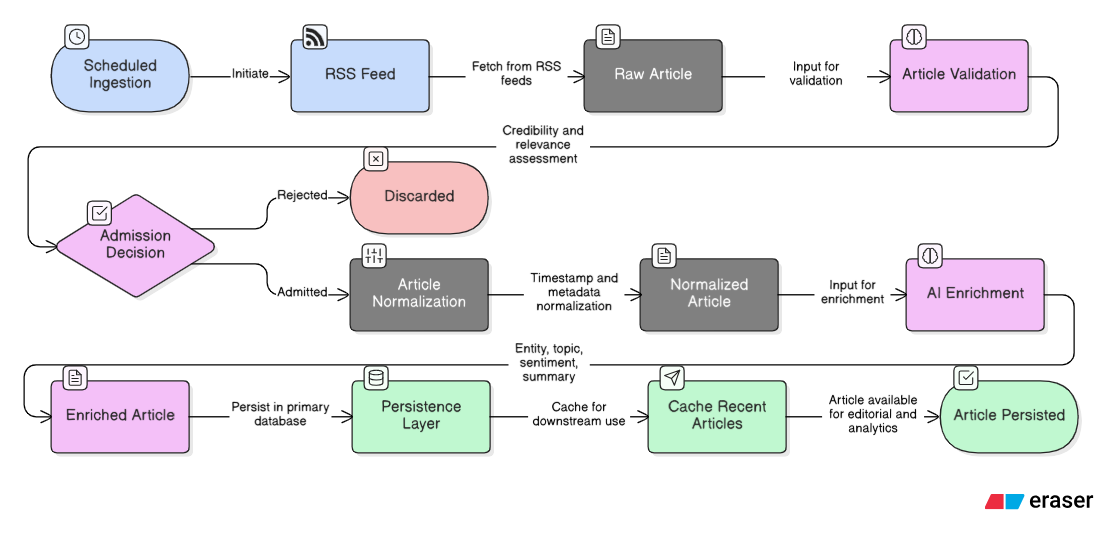}
\caption{Article ingestion and AI-based enrichment pipeline. Articles are validated, normalized, enriched, and persisted through scheduled background processing.}
\label{fig:article_pipeline}
\end{figure}

Figure~\ref{fig:architecture} presents the complete system architecture, while Figure~\ref{fig:article_pipeline} focuses on the article-level processing pipeline, from ingestion through enrichment and persistence. Together, these figures illustrate how individual articles are transformed into structured representations and integrated into long-lived editorial structures.

Within this architecture, artificial intelligence refers to a combination of large language models for semantic interpretation and comparison, embedding-based representations for similarity and clustering, and structured storage for maintaining editorial state. These components are coordinated through a controlled pipeline in which each model performs a specific, well-defined function rather than acting autonomously.

This design emphasizes consistency and continuity over time. Recent work on semantic stability and knowledge preservation in representation learning highlights the importance of maintaining interpretive coherence across long-running systems \cite{cao2021knowledge, lozano2024semantic}. The architecture presented here reflects this principle by ensuring that editorial structures evolve incrementally as new reporting arrives, rather than being recomputed in isolation.

\section{Retrospective Case Study}

This section presents a retrospective case study drawn from live system operation. The case study illustrates how the system organizes reporting into persistent editorial structures as articles arrive over time, and how editorial organization emerges from continuous processing rather than manual intervention.

The example focuses on story formation and accumulation, one visible component of the broader system architecture.

\subsection{Observation Window and Article Corpus}

The observation window spans three days, from January 18 to January 20, 2026. During this period, the system associated 12 articles from 7 independent news organizations with a single story cluster.

Article arrivals followed a common pattern observed across many topics: early related articles appeared sparsely, followed by a denser wave of reporting as the topic gained prominence. This progression makes it possible to observe how editorial structure develops over time as both reporting volume and source diversity increase.

\begin{table}[h]
\centering
\begin{tabular}{lccc}
\hline
Date & Articles & Unique Sources & Editorial State \\
\hline
January 18, 2026 & 3 & 3 & Related reporting identified \\
January 19, 2026 & 3 & 3 & Continued accumulation \\
January 20, 2026 & 7 & 5 & Persistent story established \\
\hline
\end{tabular}
\caption{Article and source accumulation for the case study story.}
\end{table}

\subsection{Story Formation and Editorial Persistence}

As articles arrived, the system evaluated semantic similarity, temporal proximity, and cross-source alignment to determine whether incoming reporting should be associated with an existing structure.

Once accumulated reporting exceeded a threshold of density and coherence, the system instantiated a persistent story object and retroactively associated earlier related articles. From that point forward, the story functioned as a stable editorial unit capable of accumulating additional context.

This behavior reflects a general design principle of the system: editorial structures emerge once sufficient informational density exists to support continuity. Stories persist across time and act as containers for accumulated reporting rather than being tied to individual publication moments.

\subsection{Cross-Source Aggregation and Editorial Memory}

As reporting accumulated, articles from multiple independent sources were integrated into a shared editorial structure. Each article contributed information to the story’s internal state, allowing continuity to be maintained across days and publishers.

With additional reporting, the story representation became more stable. Recurrent elements across sources reinforced shared context, while isolated details had proportionally less influence. This accumulation process enables the system to maintain editorial memory while remaining responsive to new information.

\section{Related Work}

Research on event detection, extraction, and tracking has examined how large text streams can be organized into meaningful structures \cite{chen2020history, xiang2019survey}. Subsequent work extended these ideas to streaming and social data, modeling story evolution and topic dynamics over time \cite{ahmed2011unified, peng2021streaming}.

Other work explored narrative structure and temporal visualization for understanding evolving information \cite{fisher2008narratives, rose2009describing}. More recent systems address knowledge accumulation and semantic stability in incremental and streaming settings \cite{cao2021knowledge, lozano2024semantic}.

This paper integrates these strands into a unified system architecture that supports autonomous editorial organization and computational investigation as continuous system capabilities.

\section{Conclusion}

This paper presented a system-oriented approach to autonomous editorial systems and computational investigation using artificial intelligence. By organizing and analyzing large volumes of independent reporting, the system transforms informational scale from a constraint into a source of insight.

The framework described here emerged directly from the design and operation of World Pulse Now (WPN News) and generalizes beyond news to other information-rich domains. Persistent story structures, editorial memory, and structured comparison provide a foundation for scalable understanding as public information ecosystems continue to expand.

\bibliographystyle{plain}

\end{document}